\documentclass[namedreferences]{solarphysics}
\usepackage[hyperref,optionalrh,solaromanenum]{spr-sola-addons} % For Solar Physics 
\usepackage{graphicx}                    % For eps figures, newer & more powerfull
\usepackage{color}                       % For color text: \color command
\usepackage{breakurl}                         % For breaking URLs easily trough lines
\RequirePackage{tikz}

\newcommand*\arcsec{\ensuremath{^{\prime\prime}}}

\chardef\us=`\_

\begin{document}

\begin{article}

\begin{opening}

\title{Solar prominence modelling and plasma diagnostics at ALMA wavelengths}

%%%%%%%%%%%%%%%%%%%%%%%%%%%%%%%%%%%%%%%%%%%%%%%%%%%
%% Authors Names
%
\author[addressref={gla},corref,email={a.rodger.1@research.gla.ac.uk}]{\inits{A.S.}\fnm{Andrew}~\lnm{Rodger} \orcid{0000-0003-0385-4581}}

\author[addressref={gla},email={Nicolas.Labrosse@glasgow.ac.uk}]{\inits{N.}\fnm{Nicolas}~\lnm{Labrosse} \orcid{0000-0002-4638-157X}}

%%%%%%%%%%%%%%%%%%%%%%%%%%%%%%%%%%%%%%%%%%%%%%%%%%%
%% Runningheads
%
%\runningauthor{}
%\runningtitle{}

%%%%%%%%%%%%%%%%%%%%%%%%%%%%%%%%%%%%%%%%%%%%%%%%%%%
%% Affilations 
%% id shold be the same with \author addressref value.
\address[id={gla}]{SUPA, School of Physics \& Astronomy, University of Glasgow, Glasgow G12 8QQ, Scotland}

%%%%%%%%%%%%%%%%%%%%%%%%%%%%%%%%%%%%%%%%%%%%%%%%%%%
%%% Abstract 
\begin{abstract}
Our aim is to test potential solar prominence plasma diagnostics as obtained with the new solar capability of the \emph{Atacama Large Millimeter / submillimeter Array} (ALMA).
   We investigate the thermal and plasma diagnostic potential of ALMA for solar prominences through the computation of brightness temperatures at ALMA wavelengths. 
   The brightness temperature, for a chosen line of sight, is calculated using densities of hydrogen and helium obtained from a radiative transfer code under non local thermodynamic equilibrium (NLTE) conditions, as well as the input internal parameters of the prominence model in consideration.
   Two distinct sets of prominence models were used: isothermal-isobaric fine-structure threads, and large-scale structures with radially increasing temperature distributions representing the prominence-to-corona transition region.
   We compute brightness temperatures over the range of wavelengths in which ALMA is capable of observing (0.32 -- 9.6mm), however we particularly focus on the bands available to solar observers in ALMA cycles 4 and 5, namely 2.6 -- 3.6mm (Band 3) and 1.1 -- 1.4mm (Band 6).
   We show how the computed brightness temperatures and optical thicknesses in our models vary with the plasma parameters (temperature and pressure) and the wavelength of observation. We then study how ALMA observables such as the ratio of brightness temperatures at two frequencies can be used to estimate the optical thickness and the emission measure for isothermal and non-isothermal prominences.
   From this study we conclude that, for both sets of models, ALMA  presents a strong thermal diagnostic capability, provided that the interpretation of observations is supported by the use of non-LTE simulation results.
\end{abstract}

%%%%%%%%%%%%%%%%%%%%%%%%%%%%%%%%%%%%%%%%%%%%%%%%%%%
%% Keywords
%

\end{opening}
%-------------------------------------------------

%%%%%%%%%%%%%%%%%%%%%%%%%%%%%%%%%%%%%%%%%%%%%%%%%%%
%% Sections
%
% \section{}%\label{s:?} 

%% Figure 
%
% \begin{figure} 
% \centerline{\includegraphics[width=0.5\textwidth,clip=]{<fig.eps>}}
% \caption{}%\label{fig:?}
% \end{figure}

%% Table
%
% \begin{table}
% \caption{}%\label{tbl:?}
% \begin{tabular}{}     
% \hline
% \multicolumn{2}{c}{<>}
% <data>
% \hline
% \end{tabular}
% \end{table}

\section{Introduction}\label{sec:intro}

The temperature structure of solar prominences remains an important question in solar physics. 
Prominences are cool, dense structures suspended in the hot sparse corona, and it is generally accepted that there is a transition region (PCTR) between the two regimes. 
The importance of the PCTR in prominence modelling has been discussed by \cite{Anzer1999}, and its effect on various spectral lines demonstrated by e.g. \cite{2001A&A...370..281H,Labrosse2004,2010SSRv..151..243L,2015ApJ...800L..13H}. However, the  nature of the PCTR and its relationship to the prominence and prominence fine-structure is not fully understood.
To address this, an accurate and reliable temperature diagnostic, capable of resolving fine-structure, is required.  

Numerous studies of solar prominences have allowed the physical parameters of the prominence plasma to be evaluated through analysing the shape and intensity of spectral lines \citep{2010SSRv..151..243L,2014LRSP...11....1P,2015ASSL..415..131L}. 
However, the complex mechanisms involved in the spectral line and continuum formation at optical or UV wavelengths in prominences presents several issues when attempting to accurately measure any kinetic temperature distribution.
Optically thick radiation requires complex non-LTE radiative transfer treatment making temperature estimations problematic. In addition, optically thick line profiles are susceptible to broadening effects, e.g. become non-Gaussian, masking the true thermally broadened profile.
As for optically thin radiation, any discernible temperature structure results from the cumulative effect from across the entire line-of-sight (LOS), thus losing information on any fine-structures present.
In reality, observations in optically thick lines will be affected by a mixture of these effects, as one integrates through more and more fine structures when observing in the optically thin line wings \citep{2008A&A...490..307G,Labrosse&Rodger2016}.

In radio wavelengths, temperature measurements are considerably more simple \citep[e.g.][]{2016SSRv..200....1W, Loukitcheva2004, 2015SoPh..290.1981H}.
In the solar millimetre/submillimetre domain the dominant emission mechanism is free-free collisional processes. 
The source function hence results from local thermodynamic equilibrium (LTE) conditions and is thus Planckian.
In the Rayleigh-Jeans domain this means the source function then increases linearly with kinetic temperature.
This causes the peak contribution function of the continuum radiation to be highly dependent on the local temperature, leading to the often used term \emph{linear thermometer}.
In \cite{2015A&A..575A..15L}, the authors conclude that, for chromospheric radiation,  brightness temperatures at millimetre wavelengths provide a reasonable measure of the thermal structure, up to resolutions of 1\arcsec.
In the context of solar flare models, \cite{2012SoPh..277..31H} synthesized the thermal continua from the optical to the mm radio, demonstrating how these continua are formed and again showing the close correspondence between brightness temperature and the kinetic temperature. 

Despite the clear advantages presented with measurements at mm radio wavelengths, observations of solar prominences have been so far limited by low spatial resolutions 
\citep[e.g.][]{1992SoPh..137..67V,1993ApJ..418..510B,1993A&A..274L..9H,1995SoPh..156..363I}. 
Studies of prominences and filaments at centimetric wavelengths have also been conducted \citep[see e.g.][]{ChiuderiDrago2001}.
Now, with the use of the \emph{Atacama Large Millimeter / Submillimeter Array} \citep[ALMA]{2011SoPh..268..165K},  solar observations with unprecedentedly high spatial resolution may be obtained.
A review of ALMA's potential contribution to solar physics is presented in \cite{2016SSRv..200....1W}.

ALMA shall offer the opportunity of a new approach to the observation and study of solar prominences. 
It is thus important to understand how we expect prominences and prominence fine structure to appear in brightness temperature when observed in ALMA's  wavelength range.
Once prominence observations are obtained it will also be important to understand how to use such measurements to infer information about the temperature and plasma structures in question.  

A study into how prominences may appear as viewed through ALMA was conducted by \cite{2015SoPh..290.1981H}.
This was done by taking an $\mathrm{H\alpha}$ coronographic image and, using the empirical relation between $\mathrm{H\alpha}$ intensity and emission measure, estimating the brightness temperature for such a plasma.
These brightness temperatures were tested using the \emph{Common Astronomy Software Applications} (CASA) package, to simulate ALMA observations.
Assumptions were however required including the use of a simple temperature structure for the prominence, whilst the simulated ALMA observations were restricted by the resolution of the instrument used to create the original $\mathrm{H\alpha}$ observation.

Simulated observations of whole prominences in the mm domain have been created by \cite{2016ApJ..833..141G} using a 3D whole-prominence fine structure model. 
The prominence fine structures are formed within dips in a synthetic prominence magnetic field. 
From the material within the fine structure the hydrogen free-free extinction coefficient and thus the brightness temperature are calculated.
This model is used to visualize the brightness temperature and optical thicknesses of prominences on the limb and on-disk filaments at a range of ALMA  wavelengths.
%The study shows how prominences may be expected to appear when viewed with ALMA, depending on whether one looks at on-disk or off-limb objects, and on the wavelength of observation.
The authors also underline the requirement for mm observations in both optically thin and optically thick wavelengths for observations of filaments, in order to distinguish between sparse, low-emitting material and dense high-absorbing material. 

In this study we test the diagnostic potential of ALMA for solar prominences by computing brightness temperatures using the 2D cylindrical solar prominence models of \cite{2009A&A..503..663G}. 
We consider two specific sets of prominence models: isothermal and isobaric fine structures, and non-isothermal large-scale structures.
These sets of models have been designed to replicate different prominence  descriptions. 
The isothermal-isobaric fine-structure models correspond to individual threads of varying temperature or pressure, whilst the large-scale non-isothermal cases describe a prominence with a cool thread core surrounded by a sheath of increasingly hot PCTR material. 
For each of these cases, we investigate the temperature and plasma diagnostic capability of the brightness temperature measurements. 

In section~\ref{sec:Modelling} we provide a description of the millimetre/sub-millimetre prominence models used in this study. 
Section~\ref{sec:diagnostics} presents brightness temperatures for both isothermal-isobaric fine-structure and non-isothermal large-scale prominence models, discussing their use as thermal diagnostics.
In section~\ref{sec:estimates} we investigate the brightness temperature as a diagnostic for plasma properties such as optical thickness and emission measure. 
Section~\ref{sec:Discussion} gives a discussion into the results of this study and their implications for solar observations with ALMA. 
Conclusions are presented in section~\ref{sec:conclusions}.

\section{Modelling}\label{sec:Modelling}

In this study we use the 2D cylindrical radiative transfer code, considering both hydrogen and helium, of \cite{2009A&A..503..663G}, referred to as C2D2E.
It considers a 5 level plus continuum hydrogen atom and a 33 level plus continuum helium atom, including 29 levels for He I and 4 levels for He II.
The electron density is calculated through the ionization equilibrium between hydrogen and helium.
Through iteratively solving the radiation transfer and statistical equilibrium equations, the code computes the NLTE energy level population densities of H and He, the electron density, and the specific intensities of several spectral lines and continua.

\subsection{Input parameters}

The input to C2D2E can consider a range of intrinsic thread parameters. 
Geometric variables include altitude above solar disc, inclination angle and thread diameter.
Internal thread parameters are gas pressure ($P_g$), temperature ($T$) and helium abundance ratio ($A_{He}$).
In this study we consider fine structure isothermal-isobaric prominence threads, as well as larger-scale threads with radially increasing temperature distributions.

High-resolution observations of solar prominences reveal increasing degrees of fine structure.
Some models suggest that prominences may be described as collections (or bundles) of fine-structure threads, either with individually varying temperatures, or with an overlying PCTR region \citep{1996ApJ...466..496F,2008A&A...490..307G,Labrosse&Rodger2016}. 
ALMA has the potential, with the correct array configurations, to observe with resolutions of 0.015\arcsec -- 1.4\arcsec~$\lambda_{mm}$ \citep{2002AN....323..271B}, and thus the potential capability to observe such fine-structure threads individually.  
To model prominence fine-structure we have assumed both isothermal temperature and isobaric pressure distributions. 
These assumptions have been chosen as over comparatively small distances, such as describe fine-structure observations, temperature and pressure variation could be small. 
To investigate isothermal-isobaric fine structures  we have created a 6 temperatures by 5 pressures grid of models to analyse. 
The input parameters can be seen in Table~\ref{table:Fmodels}.
The helium abundance is set to 0.1.
\begin{table}
	\caption{Parameters for isothermal-isobaric fine structure models}
	\label{table:Fmodels}
	%\begin{center}
		\begin{tabular}{c c}
			\hline\hline
			Parameter & Value \\
			\hline
			Temperature (K) & \{5000, 6000, 7000, 8000, 9000, 10000\} \\
			Pressure (dyn $\mathrm{cm^{-2}}$) & \{0.02, 0.05, 0.1, 0.3, 0.5\} \\
			Radius (km) & 250 \\ 
			\hline
		\end{tabular}
	%\end{center}
\end{table}

In observations of larger, or less well resolved structures, it might not be prudent to assume an isothermal temperature distribution. 
For these cases we consider larger scale, full-prominence width threads with distinct core and PCTR regions.
The prominence core is defined by an isothermal temperature distribution, whilst the temperature in the PCTR increases with radius.
Across the cylinder a constant gas pressure is assumed. 
In all non-isothermal models an ad-hoc temperature gradient is considered with the form:
\begin{equation}\label{eq:tempdist}
	\log{T(r)} = \log{T_0} + (\log{T_1} -\log{T_0}) \frac{r-r_0}{r_1-r_0} \ .
\end{equation}
$T_0$ and $T_1$ are the temperatures of the thread core and surrounding corona respectively.
The inner radius of the transition region is defined by $r_0$ whilst the radius of the cylinder is $r_1$.
This temperature distribution is not generated from any theoretical model and simply serves the purpose of showing the effect of a radial temperature gradient.
Table \ref{table:pmodels} gives the parameters used to define this set of models.
The helium abundance is again fixed at 0.1.
\begin{table}
	\caption{Parameters for large-scale, non-isothermal models}
	\label{table:pmodels}
	%\begin{center}
		\begin{tabular}{c c}
			\hline\hline
			Parameter & Value \\
			\hline
			Temperature (K) & $T_0 = 6 \times 10^3$, $T_1 = 1 \times 10^5$ \\
			Pressure (dyn $\mathrm{cm^{-2}}$) & \{0.02, 0.03, 0.05, 0.1, 0.2, 0.3, 0.5\} \\
			Inner radius (km) & 500 \\
			Outer radius (km) & 1000 \\
			\hline
		\end{tabular}
	%\end{center}
\end{table}

\subsection{mm/sub-mm continuum formation in solar prominences}

In computing the radiation at a given wavelength, the most import aspect to consider is the emission mechanisms. 
ALMA will take observations in the wavelength range between 0.3~mm and 9.0~mm \citep{2011SoPh..268..165K}. However, initial solar observations in Cycle 4 are limited to 2.6 -- 3.6~mm (Band 3) and 1.1 -- 1.4~mm (Band 6).
At millimetre/sub-millimetre wavelengths the Sun's radiation is dominated by the free-free thermal continuum \citep{2016SSRv..200....1W}.
Free-free processes are fully collisional so the continuum radiation is formed under local thermodynamic equilibrium (LTE) conditions.
The respective source function is thus given by the Planck function, assuming the velocity distribution is Maxwellian.
Using the Rayleigh-Jeans Law the continuum source function $S_\nu$ is thus:
\begin{equation}\label{eq:Rayleigh_Jeans}
	S_{\nu} = \frac{2\nu^{2}k_{B}T}{c^2} \ ,
\end{equation}
where $\nu$ is the frequency, $T$ is the temperature and $k_B$ and $c$ are the Boltzmann constant and the speed of light in a vacuum respectively.

The continuum intensity emitted in LTE over a given optical depth at frequency $\nu$ is described by:
\begin{equation}\label{eq:intensity}
	I_{\nu} = \int S_{\nu} \mathrm{e}^{-\tau_{\nu}}d\tau_{\nu} = \int \kappa_{\nu} S_{\nu} \mathrm{e}^{-\tau_{\nu}} \mathrm{d}s \ ,
\end{equation}
where $I_{\nu}$ is the specific intensity, $\tau_{\nu}$ is the optical depth and $\kappa_{\nu}$ is the monochromatic extinction coefficient per unit path length.
In the Rayleigh-Jeans limit the equation for specific intensity can be simplified:
\begin{equation}\label{eq:intensity_RJ} 
	I_{\nu} = \frac{2\nu^{2}k_{B}T_B}{c^2} \ .
\end{equation}
Here $T_B$ is the brightness temperature, i.e. the temperature a black body in thermal equilibrium with its surroundings would have if it were to emit with the same intensity, $I_{\nu}$.
Through simple comparison between equations (\ref{eq:Rayleigh_Jeans}), (\ref{eq:intensity}) and (\ref{eq:intensity_RJ}) an expression for the observable brightness temperature in terms of the kinetic temperature (usually taken as the electron temperature), the local extinction coefficient and the optical depth can be derived:
\begin{equation}\label{eq:TB}
	T_B = \int \kappa_{\nu} T \mathrm{e}^{-\tau_{\nu}} \mathrm{d}s \ .
\end{equation}
$ds$ describes an interval along a path in the LOS.
Using the known temperature distribution of the model and by calculating the position- and wavelength-dependent extinction coefficient across the thread, the observable brightness temperature is calculated. 

\subsubsection{Calculating the extinction coefficient}\label{sec:kappa}

The largest contributions to extinction in millimetre/submillimetre wavelengths are free-free extinction due to inverse thermal bremsstrahlung from ionized hydrogen and helium, and to a lesser degree $\mathrm{H}^{-}$ extinction.
Inverse thermal bremsstrahlung describes the case where an electron, in the Coulomb field of an ion, becomes excited through the absorption of a photon from the radiation field.
In cgs units the extinction coefficient due to inverse thermal bremsstrahlung, $\kappa^{ff}$, is:
\begin{equation}\label{eq:inverse_thermal_bremsstrahlung}
	\kappa^{ff}_{ion} \approx 9.78 \times 10^{-3} \frac{n_e}{\nu^{2}T^{\frac{3}{2}}} \sum_{i} Z_i^{2} n_i \times (17.9 + \ln T^{\frac{3}{2}} - \ln{\nu})  \ ,
\end{equation}
where $T$ is the temperature, $\nu$ is the frequency and $n_e$ is the electron density.
$i$ represents each ion species considered, e.g. hydrogen or helium ions.
$n_i$ and $Z_i$ are the ion density and ion charge respectively \citep{2016SSRv..200....1W}.
Equation (\ref{eq:inverse_thermal_bremsstrahlung}) is evaluated from the semi-classical inverse thermal bremsstrahlung absorption, in the absence of magnetic field, given in \cite{1985ARA&A..23..169D}.

Continuous $\mathrm{H}^{-}$ absorption occurs from two sources, photo-detachment and free-free transitions, described in equations (\ref{eq:photodetachment}) and (\ref{eq:ffHminus}) respectively:
\begin{equation}\label{eq:photodetachment}
	h\nu + \mathrm{H}^{-} \rightarrow \mathrm{H} +\mathrm{e}^{-} \ ,
\end{equation}
\begin{equation}\label{eq:ffHminus}
	h\nu + \mathrm{e}^{-} + \mathrm{H} \rightarrow \mathrm{H} + \mathrm{e}^{-*} \ .
\end{equation}
%MAYBE WRITE EQUATIONS FOR HMINUS ABSORPTION COEFFICIENTS.
The analytical expressions describing the absorption coefficients representing these two absorption mechanisms are given in \cite{1988A&A...193..189J}.
The total absorption coefficient from $\mathrm{H}^{-}$ absorption is considerably lower than that for inverse thermal bremsstrahlung, however at high temperatures its significance does increase.
Other absorption mechanisms that have been considered include Thomson and Rayleigh scattering.

\subsection{Geometry and integration method}\label{sec:Geometry}

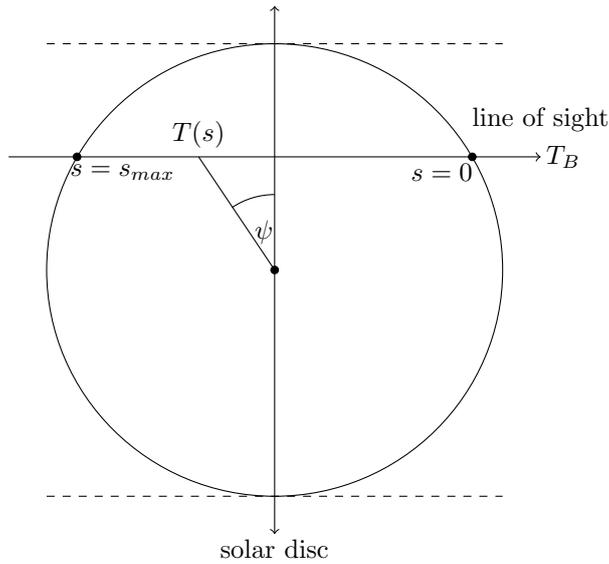
\begin{figure}
	\begin{center}
		\begin{tikzpicture}
			\draw(0,0)circle(3);
			\draw[<->](0,-3.5) -- (0,3.5);
			\draw(0,-3.7) node {solar disc};
			\draw(3.5,2) node {line of sight};
			\draw[->](-3.5,1.5) -- (3.5,1.5);
			\draw [fill=black] (-2.6,1.5) circle (0.05);
			\draw [fill=black] (2.6,1.5) circle (0.05);
			\draw [fill=black] (0,0) circle (0.05);
			\draw[dashed] (-3,-3) -- (3,-3);
			\draw[dashed] (-3,3) -- (3,3);
			\draw[-](0,0) -- (-1,1.5);
			\draw(3.8,1.5) node {$T_B$};
			\draw(-1,1.8) node {$T(s)$};
			\draw(-2.,1.3) node {$s=s_{max}$};
			\draw(2.2,1.3) node {$s=0$};
			\draw(-0.14,0.5) node {$\psi$};
			\draw(0,1) arc (270:304:-1);
		\end{tikzpicture}
	\end{center}
	\caption{Schematic diagram showing integration direction along line of sight. $s=0$ and $s=s_{max}$ correspond to the start and the end of the path of light, respectively. The dashed lines correspond to the edges of the field of view.}
	\label{fig:geometry}
\end{figure}

Whilst C2D2E can consider any range of line-of-sight directions or prominence inclinations, this paper shall only present results from off-limb threads. 
The thread is thus orientated horizontally with respect to the solar surface, whilst the line of sight of the 'observer' crosses the cylindrical axis perpendicularly. 
The altitude of all model prominences in this study is 10000~km.
A vertical field of view (FOV) is defined such that the centre of the FOV corresponds with the cylindrical axis of the thread. 
For each position in the FOV the maximal length for a horizontal light path through the thread is defined. 
Through interpolation the local temperature and absorption coefficient are determined at every point on the path, and are then integrated in the manner described in Eq.~(\ref{eq:TB}).
The optical depth, $\tau_{\nu}$, is defined such that it is zero at the edge of the cylinder closest to the observer and maximal at the opposite end of the path.
The geometry of this set-up is visualised in Fig.~\ref{fig:geometry}.

The grid of radial and azimuthal positions defined for each thread quantity, e.g. electron temperature or absorption coefficient, is finite and hence when taking values at any given position along a path will require interpolation.
The azimuthal grid has constant steps and is symmetric with respect to the ($\psi$ = 0) plane, i.e can be reduced to the range [0,$\pi$]. To interpolate across the azimuthal grid a Fourier method is utilised \citep{2005A&A..434.1165G}:
\begin{equation}\label{eq:Fourier}
	F(\psi) = \sum_{j=1}^{N_{\psi}} a_j \cos [(j - 1)\psi] \ ,
\end{equation}
where $N_{\psi}$ is the total number of positions in the azimuthal grid and $a_j$ is defined by:
\begin{equation}\label{eq:fcomponents}
	a_j = \sum_{k=1}^{N_{\psi}} B_{jk} F_{k} \ .
\end{equation}
The matrix, $B$, is defined solely by the azimuthal grid.
Fourier series interpolation has the advantages of smoothness and periodicity \citep{2005A&A..434.1165G}.

\section{Computed brightness temperatures}\label{sec:diagnostics}

\subsection{Isothermal-isobaric fine structures}\label{sec:isoT}
\begin{figure}
	\begin{center}
		\includegraphics[width=\linewidth]{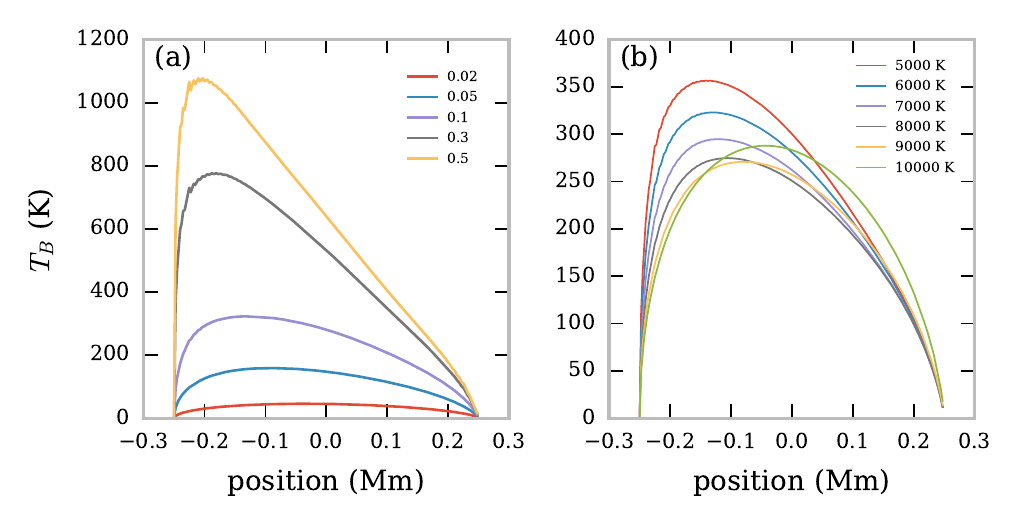}
		\caption{Computed brightness temperature across FOV for ALMA band 6, $\lambda$ = 1.3~mm.
			(a) shows the effect of increasing gas pressure (dyn $\mathrm{cm}^{-2}$) on models with a temperature of 6000~K.
			(b) shows the effect of increasing temperature (K) on models with a gas pressure of 0.1 dyn $\mathrm{cm}^{-2}$.}
		\label{fig:IsoT_Xsect}
	\end{center}
\end{figure}

In Fig.~\ref{fig:IsoT_Xsect} we show the computed brightness temperature of 1.3~mm emission (ALMA band 6) across the field-of-view (FOV) for a set of isothermal-isobaric fine-structure models. 
The FOV is orientated such that the position axis increases with increasing distance from the solar surface. 
Figure \ref{fig:IsoT_Xsect} (a) shows the brightness temperature across the field of view for models with differing isobaric pressures.
From the equation of state, low pressures reduce the overall density of the prominence resulting in a lower brightness temperature. 
The low density plasma allows ionizing incident radiation to penetrate further through the thread, creating a symmetrical brightness temperature distribution. 
When considering increasingly high pressures, the density will increase and thus too the brightness temperature. 
High density threads prevent incident radiation penetrating through the entire thread.
This causes a relative increase of ionization towards the threads lower boundary, which receives more radiation from the solar disc compared to the upper boundary. 
This in turn increases inverse-thermal bremsstrahlung absorption in this area.
The higher absorption coefficient leads to higher brightness temperatures, creating an asymmetric distribution. 

Figure \ref{fig:IsoT_Xsect} (b) shows the brightness temperature across the field of view for models with different isothermal temperatures. 
At low temperatures, an increase in the kinetic temperature causes a decrease in the brightness temperatures across the FOV. 
This will be partly due to the decreased density, through the ideal gas law, and partly due to the inverse-thermal bremsstrahlung extinction's relation with temperature (Eq.~\ref{eq:inverse_thermal_bremsstrahlung}).
At high temperatures the increase in temperature leads to further ionization of the neutral material, increasing inverse-thermal bremsstrahlung opacity, and thus the brighness temperature. 
When the material is ionized due to an overall increase in temperature, the ionizing incident radiation has a less significant effect, creating a symmetrical brightness temperature distribution across the FOV. 

\begin{figure*}
	\begin{center}
		\includegraphics[trim={0.5cm 1cm 2cm 0.5cm},clip,width=\linewidth]{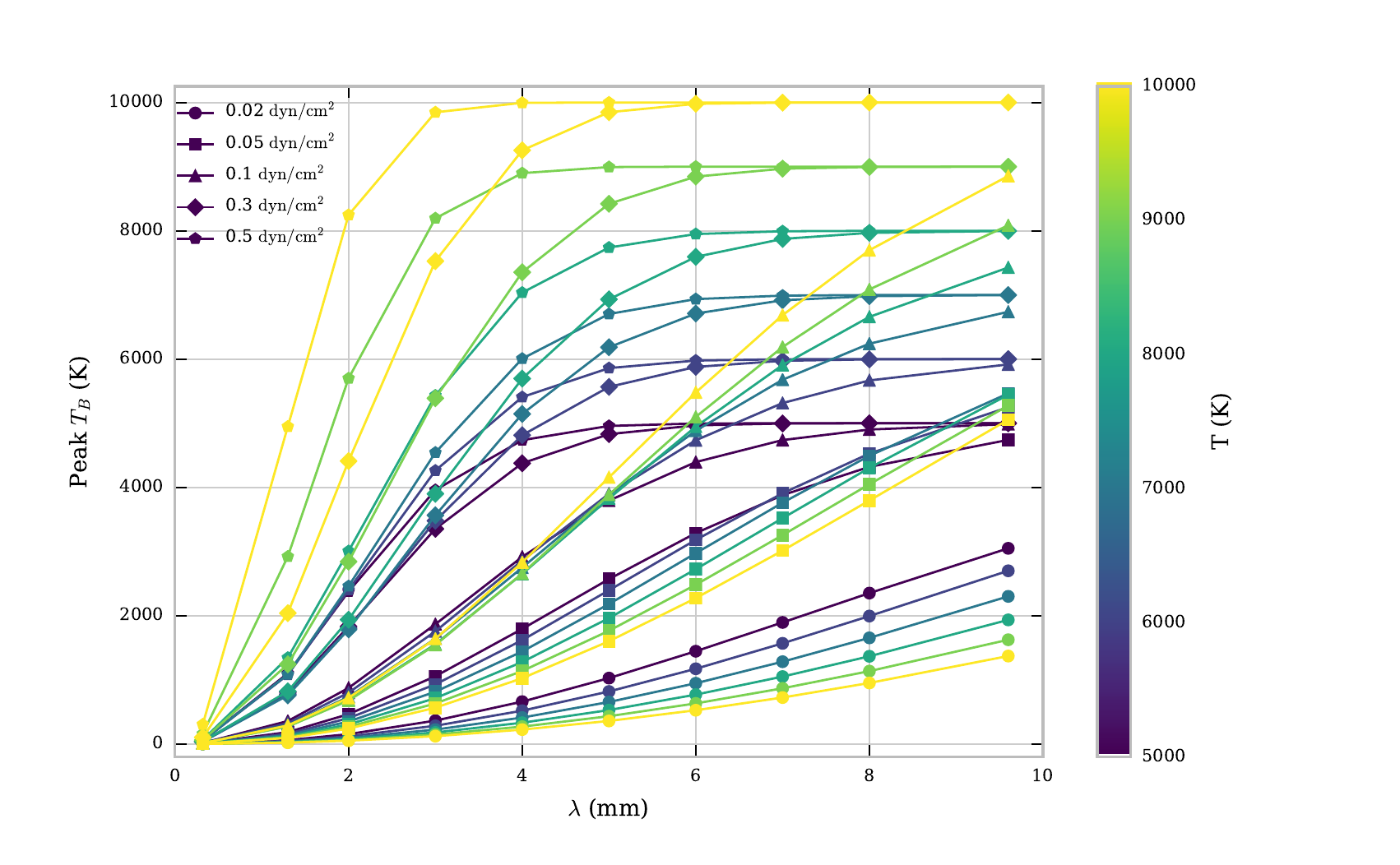}
		\caption{Relationship between peak brightness temperature and wavelength for a set of isothermal-isobaric fine structure models.
			Each colour corresponds to an isothermal temperature (K), as described in the colourbar to the right.
			Each pressure corresponds to an isobaric pressure (dyn $\mathrm{cm}^{-2}$), as described in the legend.}
		\label{fig:lambdavsTb}
	\end{center}
\end{figure*}

We show in Fig.~\ref{fig:lambdavsTb} the wavelength, temperature and pressure dependence of the peak brightness temperature in the FOV. 
The peak brightness temperature will also be dependent on the radius, i.e. length of the LOS, and the altitude of the prominence fine structure. 
Here we consider fixed values for both these quantities, whilst observationally these values could be fairly easily constrained.
From Fig.~\ref{fig:lambdavsTb} it can be seen that the peak brightness temperature increases with wavelength, until the point in which it reaches the temperature of the plasma.This should occur when the optical thickness of the observing wavelength reaches or exceeds unity.  

The brightness temperature generally increases with wavelength due to the enhanced extinction from inverse-thermal bremsstrahlung (Eq.~\ref{eq:inverse_thermal_bremsstrahlung}).
The point at which the peak brightness temperature reaches saturation with the kinetic temperature is defined by the radiation's extinction coefficient, i.e. the higher the extinction coefficient, the lower the wavelength required to reach saturation with the kinetic temperature.
The extinction coefficient is not purely wavelength dependent, but also depends on ion density and temperature: the higher the pressure, the higher the  density, which  leads to more extinction and thus higher brightness temperatures.
Increasing temperature in high pressure models leads to more ionization and thus higher brightness temperatures. 
At low pressures however, increasing the temperature decreases the overall density more than it increases the ionization, causing a decrease in extinction and brightness temperature. 

The observed optically thick mm continuum emission is most representative of the plasma near the position where the optical depth reaches unity. 
For a fully isothermal thread this measurement is thus an accurate representation of the kinetic temperature over the entire thread. 
Hence multiple optically thick wavelength observations will reproduce the same brightness temperature measurement, seen as the saturation features in Fig.~\ref{fig:lambdavsTb}.

Due to their limited spatial extent, individual observed fine-structure threads will naturally tend towards being optically thin in all bar the highest wavelengths and extinction. 
In isobaric-isothermal models with fine structure of the scale of these threads, the peak optical thickness of band 6 radiation fails to reach $\tau = 1$ for all models, whilst the optical thickness of band 3 radiation exceeds $\tau = 1$ for models at pressure of 0.3 or $0.5\,\mathrm{dyn\,cm}^{-2}$ (Table \ref{table:Fmodels}).
Increasing the size of the thread will increase the optical thickness at the observed wavelength. 

Each individual isobaric-isothermal model produces a distinct peak brightness temperature versus wavelength curve. 
If geometrical variables such as altitude or LOS width can be constrained, a brightness temperature observation of an isobaric-isothermal fine-structure thread, at known wavelength, could be used in conjunction with our set of models to set constraints on the pressure and temperature of the structure in consideration. 
If multiple observations in different wavelength bands are available, the constraints on the isobaric-isothermal model should improve greatly. 

\subsection{Non-isothermal large-scale structures}\label{sec:non-isoT}

\begin{figure}
	\begin{center}
		\includegraphics[width=\linewidth]{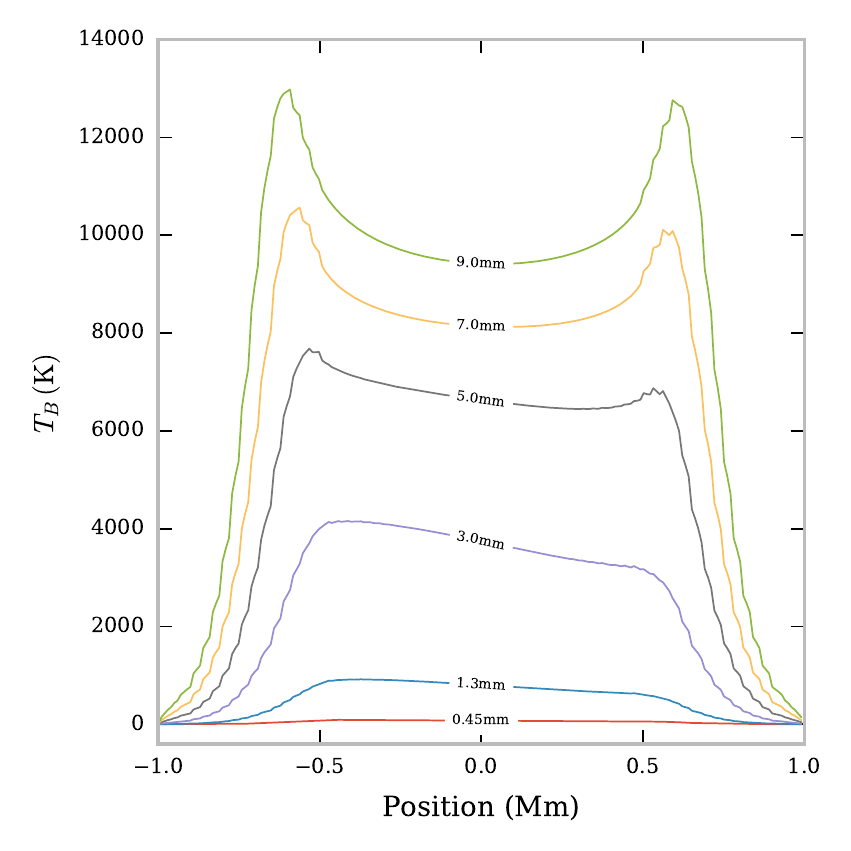}
		\caption{Variation of brightness temperatures across FOV in non-isothermal large-scale prominence models. 
			Isobaric pressure is 0.1~dyn cm$^{-2}$ and FOV is orientated vertically in the solar atmosphere with 
		the positive x-axis directed radially away from the Sun.}
		\label{fig:p4_Tb}
	\end{center}
\end{figure}
\begin{figure}
	\begin{center}
		\includegraphics[width=\linewidth]{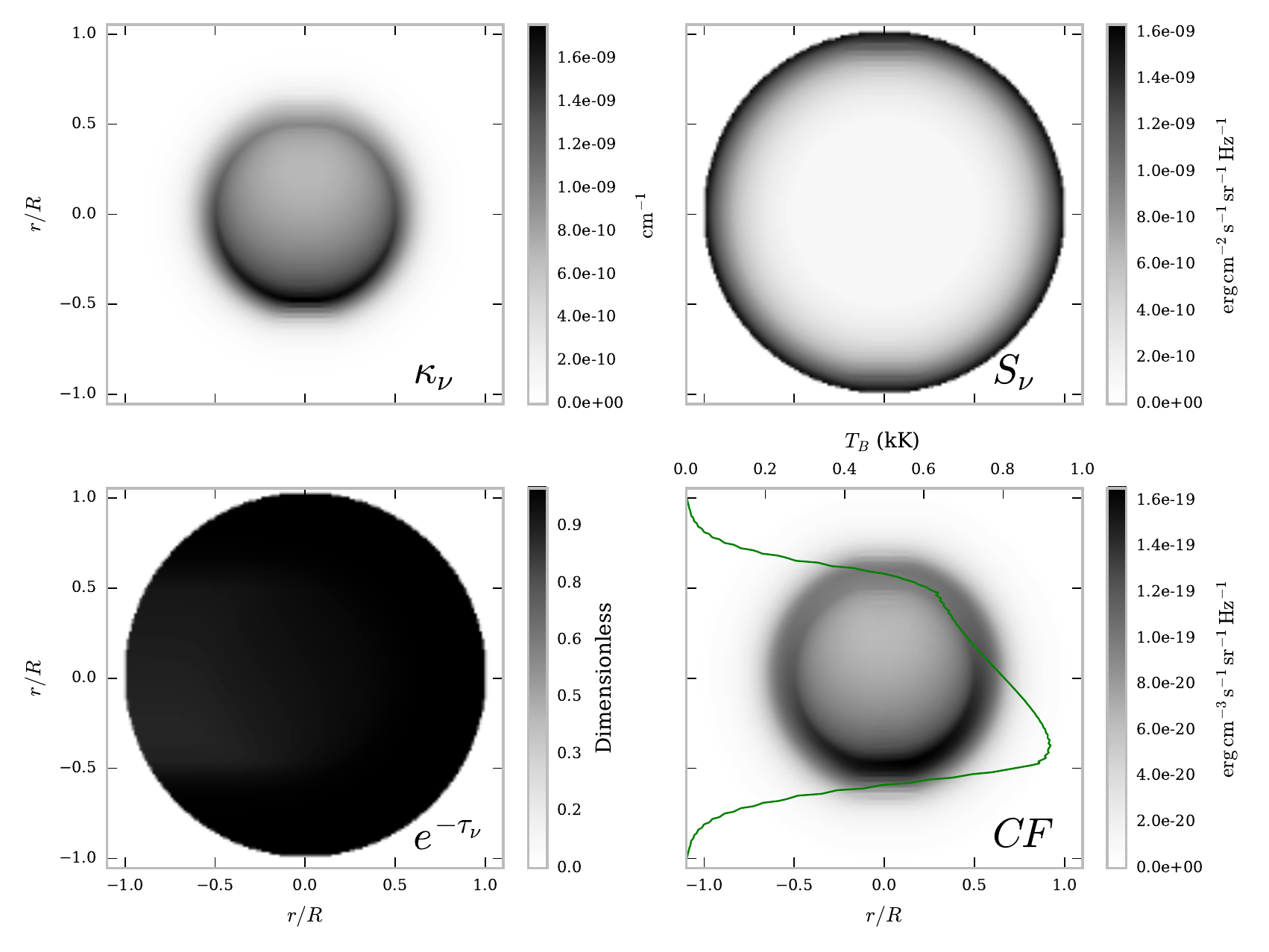}
		\caption{Brightness temperature FOV for non-isothermal large-scale prominences with gas pressure of 0.1 dyn $\mathrm{cm}^{-2}$ at $\lambda = 1.3$~mm. The \emph{top left} figure shows a map of the absorption coefficient, the \emph{top right} shows the source function and the \emph{bottom left} shows the optical thickness attenuation term. The resulting contribution function map is in the \emph{bottom right} hand panel. Integrating over each horizontal line of sight results in the brightness temperature (K) curve, \textcolor{green}{solid green line}.}
		\label{fig:formation1.3}
	\end{center}
\end{figure}
\begin{figure}
	\begin{center}
		\includegraphics[width=\linewidth]{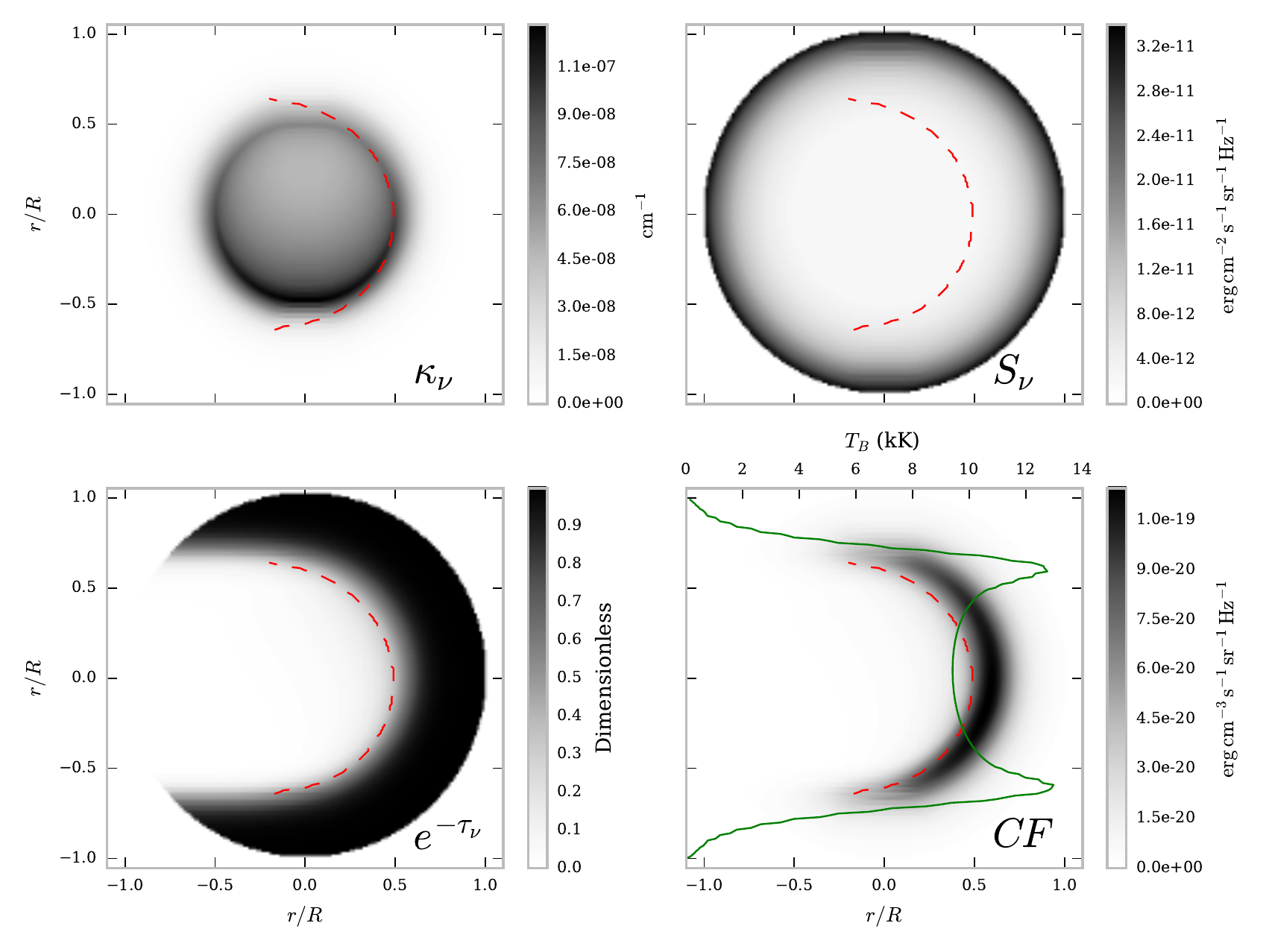}
		\caption{Brightness temperature FOV for non-isothermal large-scale prominences with gas pressure of 0.1 dyn cm$^{-2}$ at $\lambda = 9$~mm. The \emph{top left} figure shows a map of the absorption coefficient, the \emph{top right} shows the source function and the \emph{bottom left} shows the optical thickness attenuation term. The resulting contribution function map is in the \emph{bottom right} hand panel.
		Integrating over each horizontal line of sight results in the brightness temperature (K) curve, \textcolor{green}{solid green line}.
		The \textcolor{red}{dashed red line} shows the $\tau$ = 1 line.}
		\label{fig:formation9.0}
	\end{center}
\end{figure}

In Fig.~\ref{fig:p4_Tb} we show the variation of brightness temperature across the FOV for a large-scale prominence structure with a radially increasing temperature and a pressure of 0.1~$\mathrm{dyn\,cm^{-2}}$ (Table \ref{table:pmodels}) at several mm/sub-mm wavelengths. 
Immediately it can be seen that there are two regimes that can describe the brightness temperature variation. The emission at  
0.45, 1.3 \& 3.0~mm is optically thin ($\tau < 1$) in this model, and thus displays a smooth, asymmetric variation across the FOV. Emission at 
5.0, 7.0 \& 9.0~mm is optically thick ($\tau \geq 1 $) and shows a nearly symmetric, dual-peaked variation.
The formation of these two regimes is better understood through considering the contribution function and its constituent parts (see Figs~\ref{fig:formation1.3} and \ref{fig:formation9.0}).

The formation plots show how the absorption coefficient, source function and optical thickness attenuation terms combine across the thread to produce a map of the contribution function, seen in bottom right panel of each figure. 
Figure \ref{fig:formation1.3} shows the optically thin case where the attenuation, $e^{-\tau_{\nu}}$, is close to 1 and nearly uniform across all LOS in this cross-section. 
Photons at mm wavelengths thus travel through the thread mostly unperturbed allowing the plasma at the far side of the LOS to have almost equal contribution as the material near the surface closest to the observer.  
The prominence plasma at UV wavelengths is however non-transparent thus leading to an increase in ionizing radiation incident on the lower side of the thread. 
This leads to higher ionization and therefore higher contribution function at the side of the thread closer to the solar disc. 
Integrating the contribution function along each horizontal path in the FOV results in the brightness temperature curve as seen in the bottom right figure. 
The temperature variation is azimuthally symmetrical, hence so too is the source function (Eq.~\ref{eq:Rayleigh_Jeans}). 

Figure \ref{fig:formation9.0} is an example of a predominantly optically thick case, where within the central part of the thread the attenuation term has a large effect. 
The red-dashed line represents the $\tau$ = 1 line, i.e. the point in which the thread becomes optically thick. 
The high attenuation within the central region leads to a crescent shaped contribution function map, around the $\tau$ = 1 line. 
The core and far side of the thread are thus under-represented in the integration over the LOS. 
Each of the two peaks in the brightness temperature variation correspond to the extremal heights for which the thread is optically thick. 
This is due to a longer LOS intersecting through more high temperature, PCTR material. 
Beyond these points the plasma is once again optically thin and the brightness temperature drops off quickly. 

The incident radiation ionizing the optically thick plasma leads to an increase in absorption coefficient but also an increase in attenuation from the $e^{-\tau_{\nu}}$ term. 
This produces an almost symmetrical brightness temperature variation. 

\subsubsection{Thermal diagnostic}

\begin{figure}
	\begin{center}
		\includegraphics[width=\linewidth]{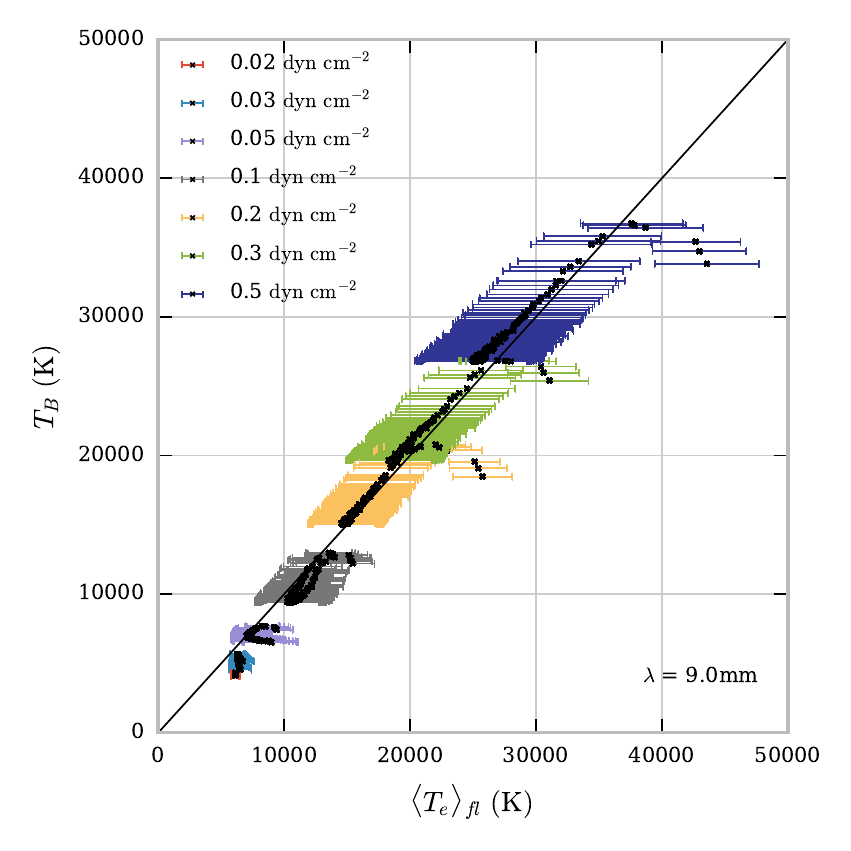}
		\caption{Relationship between brightness temperature and average kinetic temperature over the formation layer in the LOS. The formation layer is defined to be the region with 0.7 times the maximum contribution function or greater, for each LOS in which the plasma is optically thick. The error bars show the standard error in the mean for the mean kinetic temperature. Each colour corresponds to a different pressure as defined in the legend.}
		\label{fig:tempdiagnostic9.0}
	\end{center}
\end{figure}

It is difficult to determine a temperature distribution from brightness temperature measurements of optically thin plasma.
The resultant brightness temperature will be an integration over potentially large temperature variations, hence losing any discernible structure. 
Conversely, optically thick emission is representative of a specific formation region, i.e. the Eddington-Barbier approximation states $T_B(0) \approx T(\tau = 1)$.

To investigate how a brightness temperature measurement relates to the prominence plasma in the formation layer we define an effective formation layer as the parts of the prominence where the contribution is equal to or greater than 70\% of the maximum contribution function for each LOS. 
The effective formation temperature ($\langle T_e \rangle_{fl}$) is then found by taking the contribution function weighted mean of the temperature distribution across the effective formation layer. 

\begin{figure}
	\begin{center}
		\includegraphics[trim={1cm 0.5cm 2cm 0.5cm},clip,width=\linewidth]{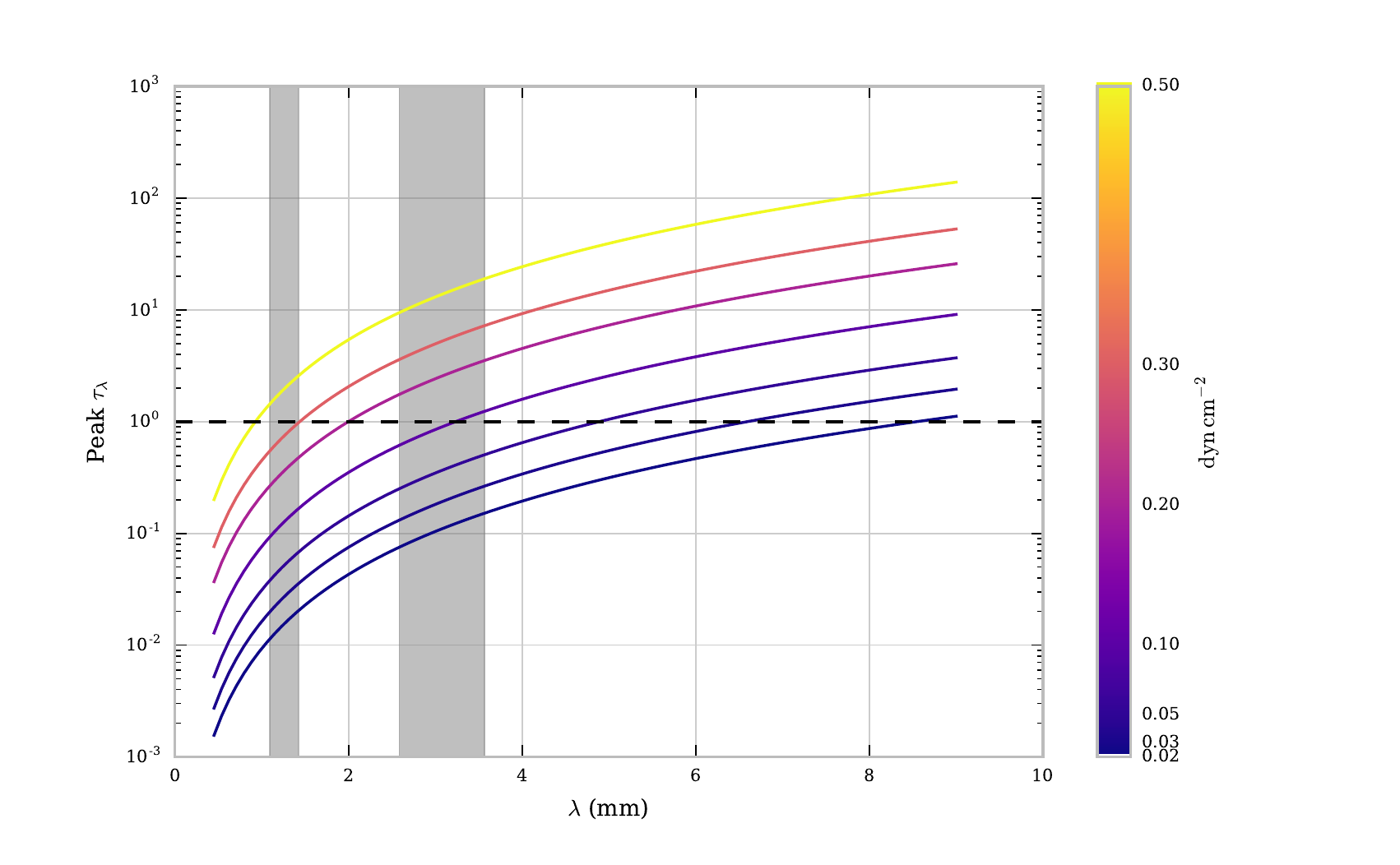}
		\caption{Relationship between peak optical thickness and wavelength for a set of non-isothermal large scale structures at various pressures.
		The dashed line represents the transition between optically thin and thick plasma.
		The two grey shaded areas cover ALMA bands 3 and 6.}
		\label{fig:tau_lambda}
	\end{center}
\end{figure}

Figure~\ref{fig:tempdiagnostic9.0} shows the relationship between computed brightness temperature and mean temperature of effective formation layer for $\lambda = 9.0$~mm, across a range of isobaric pressures. 
Each point on the graph represents an optically thick LOS.
Optically thin LOS are ignored as their contribution functions are very broad across the LOS, giving poor temperature diagnostics. 
At higher pressures more of the thread is optically thick, and hence more LOS points are shown on the graph. 
For the majority of LOSs the brightness temperature scales linearly with the mean temperature of the formation layer, with only some deviation at high temperatures in each model.
At low pressures, the effect of lower boundary ionization from incident radiation can again be seen through the splitting of the trend into two separate lines. 
Observations of brightness temperatures at optically thick wavelengths, such as $\lambda = 9.0$~mm, are thus fairly good indicators of the mean electron temperature of specific areas of the prominence. 
With high resolution observations of multiple optically thick wavelength bands, it should be possible to build up understanding of the temperature distribution within the prominence structure, as each wavelength band should be formed at a different formation layer. 
In Cycle 4 and 5 of ALMA however, the only two bands available to solar physicists are bands 3 \& 6. 
These bands are significantly less optically thick than radiation at $\lambda = 9.0$~mm, with $\tau$ only exceeding unity at the centre of the thread for models with high pressures.
%greater than 0.1 dyn~$\mathrm{cm^{-2}}$ for band 3 and 0.5 dyn~$\mathrm{cm^{-2}}$ for band 6 in any of our computations. 
The relationship between wavelength and peak optical thickness is shown in Fig.~\ref{fig:tau_lambda} for the non-isothermal models described in Table~\ref{table:pmodels}.
The two grey shaded areas represent ALMA bands 3 and 6.
As expected, the peak optical thickness (i.e. the maximum optical thickness found in each model as the line of sight is varied) increases with wavelength and with pressure.
Figure~\ref{fig:tau_lambda} shows that a structure of a similar size to what is modelled here (radius $\sim 1000$~km) observed with ALMA in bands 3 and 6 can only be expected to be optically thick at high pressures, i.e. greater than 0.5 dyn~$\mathrm{cm^{-2}}$.  

\section{Plasma diagnostics}\label{sec:estimates}

This section investigates the potential use of a brightness temperature ratio as a diagnostic for the optical thickness of the observed radiation and the emission measure of the emitting plasma.
Here we follow a similar approach to that of \cite{1993ApJ..418..510B}.
In the fourth observational cycle of ALMA, the two wavelength bands available to solar physicists are band 6 (1.3~mm) and band 3 (3.0~mm). 
The brightness temperature ratio, $R$,  is defined as: 

\begin{equation}\label{eq:Tbratio}
	R = \frac{T_{B,1.3}}{T_{B,3.0}} ,
\end{equation}
where $T_B$ is the brightness temperature and the subscripts 1.3 and 3.0 denote the wavelengths at 1.3~mm and 3.0~mm respectively. 
These subscripts will continue to be used throughout this section. 

If a constant temperature can be assumed across the prominence, Eq.~(\ref{eq:Tbratio}) can be expressed in terms of the optical thicknesses of the two measurable wavelengths, 

\begin{equation}\label{eq:R_ratio}
	R \approx \frac{T_e(1-e^{-\tau_{1.3})}}{T_e(1-e^{-\tau_{3.0}})} = \frac{1-e^{-\tau_{1.3}}}{1-e^{-\tau_{3.0}}} .
\end{equation}
In this case, the brightness temperature ratio is related to the two optical thicknesses only. 
The optical thickness, at a given wavelength $i$, can then be approximated as 
\begin{equation}\label{eq:tau_approx}
	\tau_i \approx \langle \kappa_i \rangle L,
\end{equation}
where $\kappa_i$ is the wavelength-dependent continuum absorption coefficient and $L$ is the length of the line of sight.
Using this assumption, the optical thicknesses of the two observable wavelengths are related as follows:

\begin{equation}\label{eq:opacity_ratio}
	\tau_{1.3} \approx \frac{\langle \kappa_{1.3} \rangle }{ \langle \kappa_{3.0} \rangle}\tau_{3.0} = K \tau_{3.0} , 
\end{equation}
where $K$ has been defined as the dimensionless opacity ratio. 

%\subsection{Opacity ratio at ALMA wavelengths}
As described in section \ref{sec:Modelling}, the largest source of opacity in the mm/sub-mm continuum is inverse thermal bremsstrahlung (Eq.~\ref{eq:inverse_thermal_bremsstrahlung}). 
Due to its dominance it is hence reasonable to estimate the opacity ratio whilst only considering contribution from inverse thermal bremsstrahlung.
In this case, $K$ is defined as 

\begin{equation}\label{eq:K_ratio}
	K = \frac{\nu_{3.0}^2 (17.9 + ln(T^{3/2}) - ln(\nu_{1.3}))}{\nu_{1.3}^2 (17.9 + ln(T^{3/2}) - ln(\nu_{3.0}))} .
\end{equation}
The opacity ratio is thus dependent only on the known observational frequencies and an isothermal temperature for the line of sight.

\begin{figure}
	\begin{center}
		\includegraphics[width=\linewidth]{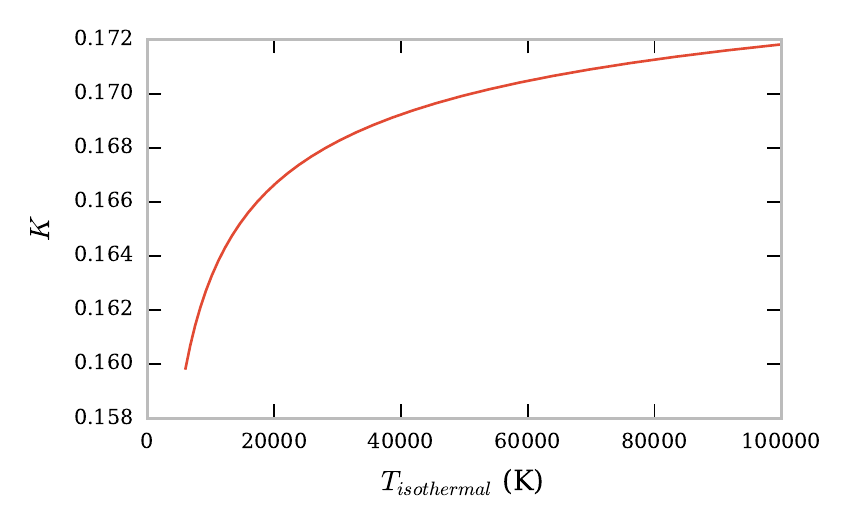}
		\caption{Variation of opacity ratio for ALMA wavelengths $\lambda_1 = 1.3$~mm and $\lambda_2 = 3.0$~mm with temperature (Eq.~\ref{eq:K_ratio}).}
		\label{fig:K_ratio}
	\end{center}
\end{figure}

To understand how the opacity ratio $K$ may vary with temperature in a prominence, it was calculated according to Eq.~(\ref{eq:K_ratio}) using wavelengths 1.3~mm and 3.0~mm, for a range of temperatures (Fig.~\ref{fig:K_ratio}), from low core temperatures of around 5000~K to extreme PCTR temperatures of $10^5$~K.  
From Fig.~\ref{fig:K_ratio}, it is clear that $K$ only varies slightly across this temperature range, and for any expected prominence temperature, $K \ll 1$ is an acceptable assumption. 
If the electron temperature of the prominence is known, or can be suitably assumed, a bound on the magnitude of the opacity ratio can be set.

\subsection{Estimating optical thickness}\label{sec:tau_est}
 
With a value for the opacity ratio, the optical thickness of either wavelength can be estimated by substituting Eq.~(\ref{eq:opacity_ratio}) into Eq.~(\ref{eq:Tbratio}).
The resulting equation can be solved analytically by expanding the exponential terms only up to the 2nd order.
This leads to:

\begin{equation}\label{eq:tau2}	
	\tau_{3.0} = \frac{2(K - R)}{K^2 - R} . 
\end{equation}
Whilst this solution is satisfactory for high temperatures and low optical thicknesses, it was found to underestimate the optical thickness as the latter increased. 
To improve the estimation for higher optical thickness cases, a numerical solution must be found instead. 
This was done by finding the roots of the function: 

\begin{equation}\label{eq:tau2_numerical}
	f(\tau_{3.0}) = \sum_{n=1}^{N} \frac{(-1)^n (K^n - R)}{n!} \tau_{3.0}^{n-1},
\end{equation}
using the Newton-Raphson method.
%The function used to do this was SciPy's \emph{scipy.optimize.newton()} \cite{SCIPY}. 
$N$ is the order to which the exponential terms are expanded. 
For the highest pressure, highest optical thickness models, this method reached a constant solution for all $N \geq 18$. 

We tested this method for optical thickness estimation using the same set of isothermal-isobaric fine-structure models as discussed in section~\ref{sec:isoT}. 
The orientation of each prominence was as described in section \ref{sec:Geometry}.
We obtained brightness temperatures at both 1.3~mm and 3.0~mm, which were then used to calculate the ratio $R$ for all points in the field of view.

\begin{figure}
      \begin{center}
	    \includegraphics[width=\linewidth]{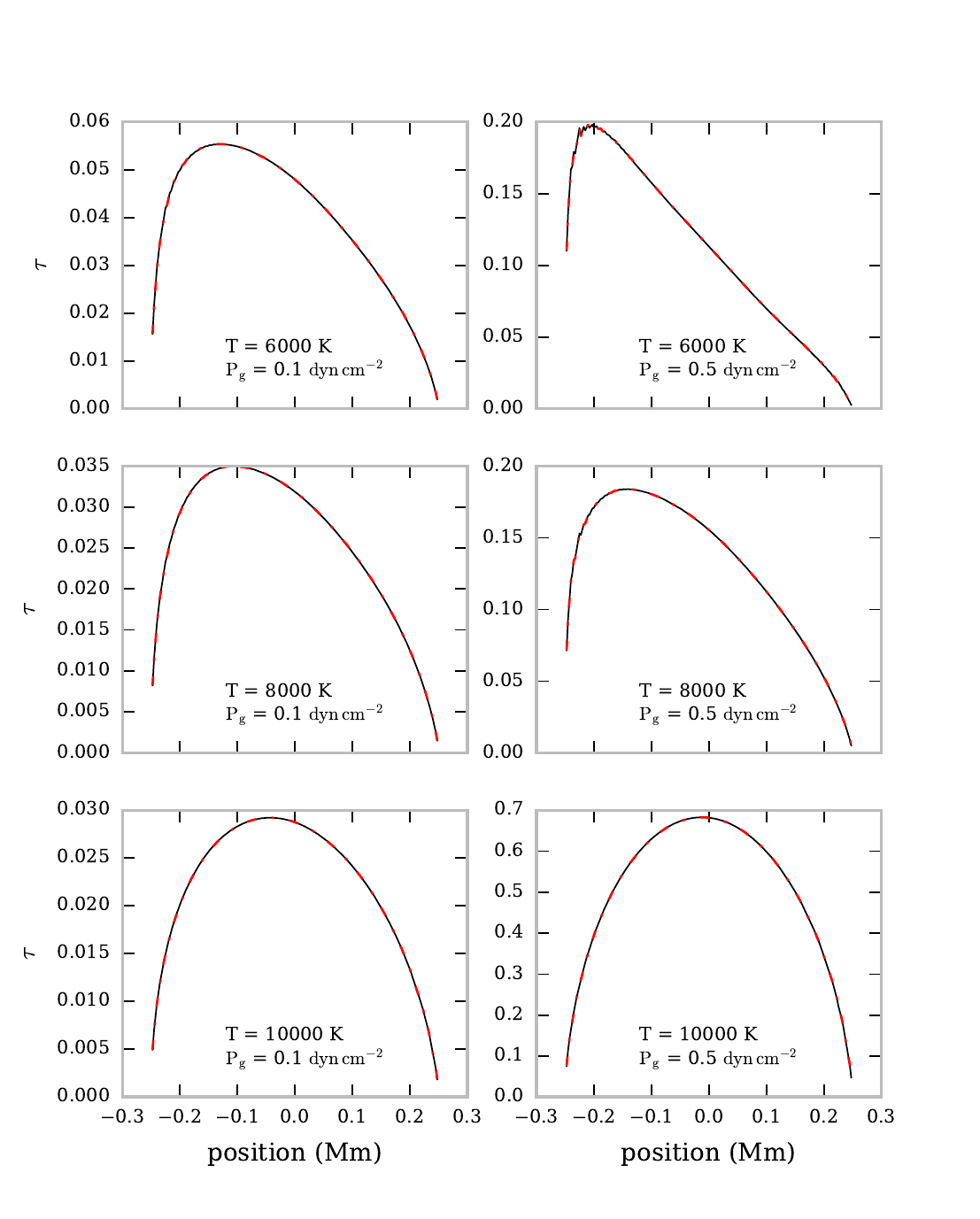}
	    \caption{Variation of estimated (\textbf{solid black line}) and computed (\textcolor{red}{dot-dashed red line}) optical thickness with the field of view, for six isothermal prominence models at $\lambda$=1.3~mm. The field of view is orientated vertically in the solar atmosphere with the positive x-axis directed radially away from the Sun.}
	    \label{fig:tau_estimates.png}
      \end{center}
\end{figure}

In Fig.~\ref{fig:tau_estimates.png} the estimated variation of the LOS optical thicknesses of 1.3~mm radiation across the FOV  is shown for a sub-set of the isothermal-isobaric models.
The opacity ratio $K$ was calculated using Eq.~(\ref{eq:K_ratio}), and the known isothermal temperature for each model. 
We find that the optical thickness estimation matches well the computed optical thicknesses across a large range of isothermal temperatures. 
The estimation is similarly as accurate for $\lambda = 3.0$~mm. 

It should be noted that our computed brightness temperatures are idealised and noiseless and that an attempt to use this method with real brightness temperature measurements would have an associated uncertainty.
This uncertainty would likely have a significant effect if both observation wavelengths are highly optically thin, i.e. $\tau \ll 1$. 
Both brightness temperatures will be low and will hence present a small SNR.
Eq~(\ref{eq:R_ratio}) can be simplified to $R \approx K(T)$ here.  

\subsection{Estimating emission measure}

Once the optical thickness at a given wavelength can be estimated sufficiently well, there is only a small step to being able to estimate the average emission measure, for a given line-of-sight.
Again, we assume that in the solar mm/sub-mm domain, opacity is greatly dominated by free-free inverse thermal bremsstrahlung.
Substituting Eq.~(\ref{eq:inverse_thermal_bremsstrahlung}) into Eq.~(\ref{eq:tau_approx}), the mean emission measure can be written as:

\begin{equation}\label{eq:EM_estimate}
      \langle EM \rangle = \frac{\tau_{\nu} \nu^2 T^{3/2}}{9.78 \times 10^{-3}(17.9 + ln(T^{3/2}) - ln(\nu))}, 
\end{equation}
where $\nu$ is the frequency of the observation and $EM$ is defined as  
\begin{equation}\label{eq:EM}
      EM = n_e \sum_j Z_j n_j L . 
\end{equation}
$n_e$ is the electron density with $Z_j$ and $n_j$ being the charge and density of ion species $j$, respectively. 

\begin{figure}
      \begin{center}
	    \includegraphics[width=\linewidth]{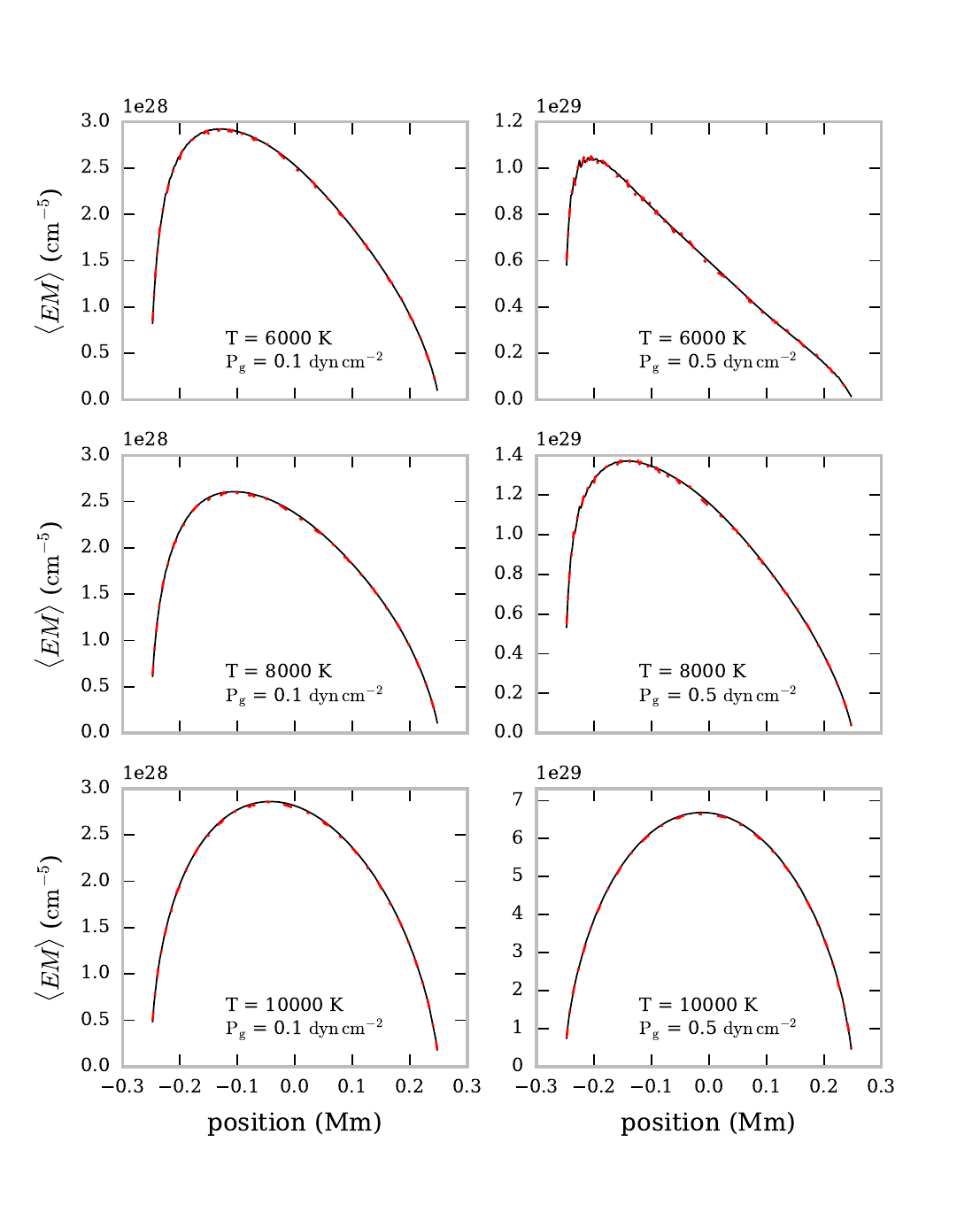}
	    \caption{Variation of estimated (\textbf{solid black line}) and computed (\textcolor{red}{dot-dashed red line}) mean emission measure with the field of view, for six isothermal prominence models. The field of view is orientated vertically in the solar atmosphere with the positive x-axis directed radially away from the Sun.}
	    \label{fig:EM_est}
      \end{center}
\end{figure}

Using the optical thickness values calculated in section \ref{sec:tau_est}, this method was used to estimate the mean emission measure, as seen in Fig.~\ref{fig:EM_est}.
The estimated value is very close to the  calculated value, with only very slight underestimation for low isothermal temperature models. 
Both 1.3~mm and 3.0~mm produce the same estimate for emission measure value. 

It has been shown that the optical thickness and the emission measure can be well estimated for isothermal prominence models with known temperatures. However, a real prominence will likely have a significantly more complex temperature distribution.
The next section discusses the effectiveness of these estimation methods on non-isothermal prominences with radially increasing temperature structures.

\subsection{Non-isothermal case}

\begin{figure}
      \begin{center}
	    \includegraphics[width=\linewidth]{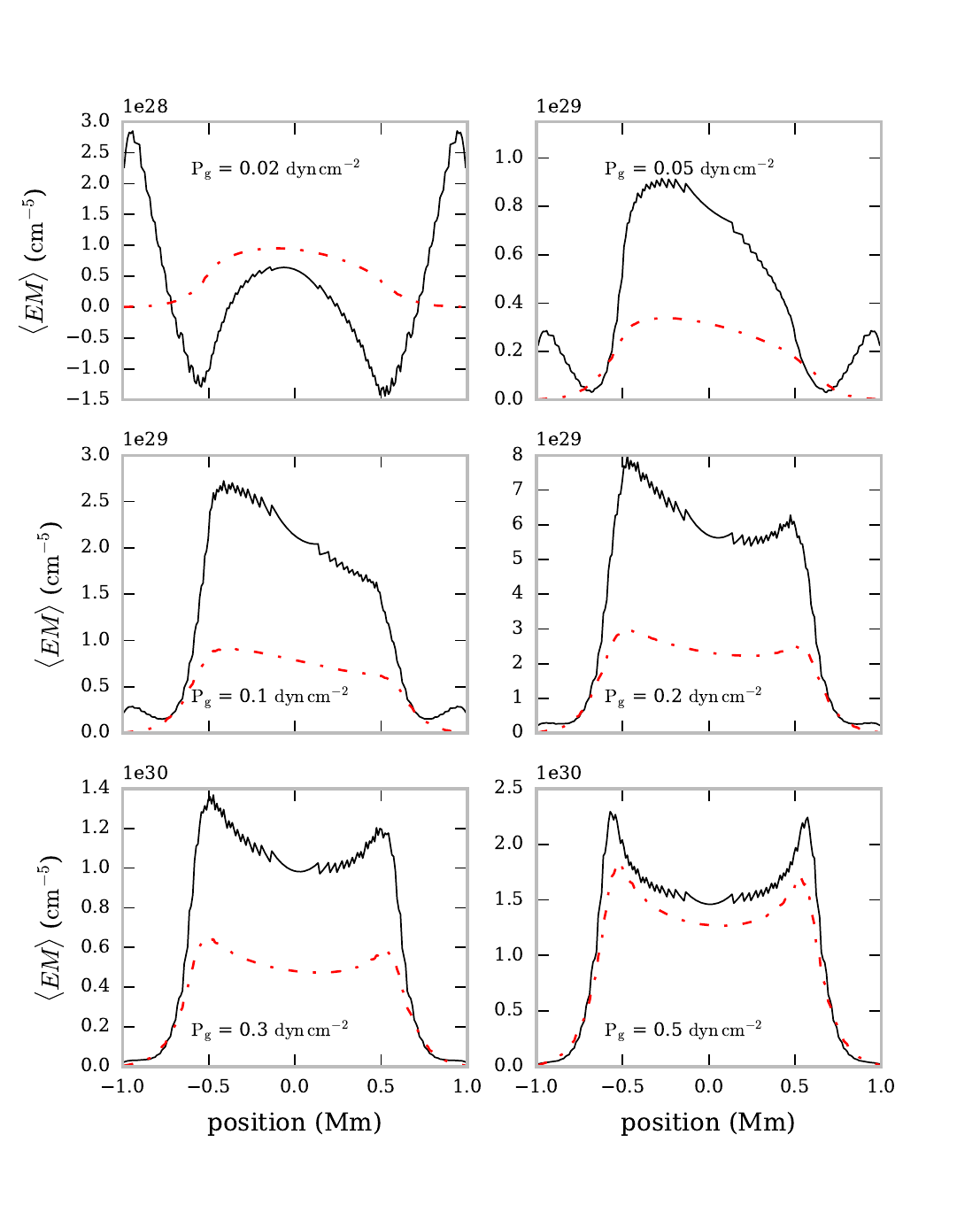}
	    \caption{Variation of estimated \textbf{solid black line}) and computed (\textcolor{red}{dot-dashed red line}) mean emission measure with the field of view, for six non-isothermal large-scale prominence models. The field of view is orientated vertically in the solar atmosphere with the positive x-axis directed radially away from the Sun.}
	    \label{fig:EM_pmodels}
      \end{center}
\end{figure}

The optical thickness and emission measure were estimated for a set of prominence models with a temperature variation increasing radially from an isothermal core (see Table~\ref{table:pmodels} and section~\ref{sec:non-isoT}). 
The brightness temperature measurements at 1.3~mm and 3.0~mm from  each model was used to estimate the respective optical thicknesses and the mean emission measure, using the method described above.
The average temperature for each LOS across the FOV was calculated and used as separate input temperature approximations. 
The average emission measure estimated for a selection of non-isothermal large-scale prominence structures estimated using this method is shown in Fig.~\ref{fig:EM_pmodels}.
The estimated values often overestimate (by up to a factor $\sim 3$) the true value of the average emission measure for each model. 

For optically thin radiation the resultant brightness temperature is a result of the integral of the contribution function across the whole LOS. 
However, due to each LOS being non-isothermal and the temperature dependence of the mm emission contribution function, different layers will present different overall contributions to the output brightness temperature. 
Hence the output brightness temperature may not be representative of the average temperature of the LOS. 
In the top left panel of Fig.~\ref{fig:EM_pmodels} the plasma is sparse enough that the estimation produces a negative value for the optical thickness and thus the emission measure. 
This is obviously unphysical and is caused by the brightness temperature ratio exceeding the opacity ratio (Eq.~\ref{eq:tau2}), as the value for the opacity ratio is created from an unrepresentative temperature estimate for the sparse material. 

In the optically thick case the emission is representative of a small region within the LOS and two wavelength observations will likely be representative of differing layers, and thus estimations using an average temperature for the LOS are difficult. 

\section{Discussion}\label{sec:Discussion}

The results of this study show clearly that whilst ALMA  will present an important opportunity to improve understanding in solar prominences, the interpretation of observed brightness temperatures may be less straightforward than what the often used phrase \emph{linear thermometer} might suggest.
There will be several decisions observers will have to make whilst choosing how to analyse a prominence observation with ALMA. 
In this paper we have outlined two contrasting methods for approaching the analysis of prominence mm/sub-mm observations, either assuming a purely isothermal structure, or accepting the requirement for a varying temperature distribution. 

The adopted approach  requires knowing how the prominence as a whole appears when viewed with ALMA, and how its appearance is altered with observation band and interferometer array configuration. 
The visualisation of how whole prominences may look when observed with ALMA has been simulated by \cite{2015SoPh..290.1981H} and \cite{2016ApJ..833..141G}.
The former used a method to convert a calibrated H$\alpha$ intensity map into an estimated brightness temperature observation. 
Due to the nature of H$\alpha$ emission, the output brightness temperature map would primarily reflect the cool prominence core material. 
The spatial resolution of the simulated brightness temperature map is also limited by the resolution of the H$\alpha$ observation.
The latter study uses a 3D whole prominence fine structure model based on dips in a model magnetic field to calculate the hydrogen free-free extinction coefficient and thus the brightness temperature, for a wavelength and LOS.
Both visualisation methods suggest that mm/sub-mm radiation could be observed from prominence fine structures with the resolution of ALMA.
If ALMA is indeed capable of resolving individual fine-structure threads separately from the prominence as a whole, such observations could be analysed by adopting an isothermal-isobaric assumption. 

In section~\ref{sec:isoT} we presented the results for brightness temperature calculations for a grid of isobaric-isothermal fine-structure prominence threads. 
For ALMA's observational cycles 4 and 5, solar  observations will have wavelength bands 3 and 6 available. 
For individual fine-structure threads, these wavelengths are likely to be optically thin, unless large pressures are present. 
As the radiation is unlikely to be optically thick, direct measurement of the kinetic temperature from a saturated brightness temperature is unlikely. 
However, non-LTE isobaric-isothermal radiative transfer models, such as we have presented in this paper, could be used alongside optically thin measurements to constrain isobaric pressure and isothermal temperature parameters. 
%If estimates for the thread radius and altitude are possible. 
Multiple wavelength observations will help to constrain the models further, although presently cycle 4 or cycle 5 will not offer simultaneous band 3 and band 6 observations. 
Measurements from different channels within each band may help improve to constrain the model, however ideally a larger wavelength spread would work better. 

If the isothermal-isobaric assumption were to be expanded to structures larger than the fine-structure threads discussed in this study, the optical thickness of band 3 and perhaps band 6 radiation may exceed $\tau = 1$, allowing direct measurement of the kinetic temperature. 
If the large prominence structure being observed is perceivably formed of a number of fine-structure threads, a multithread solution could be instead considered \citep{2008A&A...490..307G,Labrosse&Rodger2016}. 
When a number of threads are viewed along the LOS, the optical thickness at a given wavelength increases, perhaps beyond unity such that direct temperature measurements may be obtained.
This would however require the assumption to be expanded from each individual fine-structure thread being isothermal and isobaric to all threads along the LOS being mutually isothermal and isobaric. 

In section~\ref{sec:estimates} we showed how, for an isothermal prominence thread, the ratio of brightness temperature measurements at two wavelengths can be a useful plasma diagnostic for optical thickness and average emission measure along the LOS. This diagnostic requires the knowledge of the kinetic temperature of the observed plasma.
Plasma estimations using only highly optically thin measurements may be difficult as the brightness temperatures associated with the observations would be low.
%This diagnostic requires the same limitations as described for thermal diagnostics with an isothermal assumption.

Analysing prominence observations of large scale structures, or numerous collections of fine structures, a blanket isothermal assumption will likely be invalid. 
To understand a varying temperature distribution, such as is expected in the PCTR, would require accurate positional sampling of the local kinetic temperatures throughout the structure. 
Optically thin emission will likely be of little use here as the resulting brightness temperature will describe contributions from across the entire LOS, presenting at best an average kinetic temperature over what could potentially be large temperature gradients.   
If optically thick observation is possible, the resultant brightness temperature is most representative of a localised region within the LOS around the position where $\tau = 1$. 
Different wavelength ALMA bands, sub-bands and channels within sub-bands, will be representative of different regions due to the sensitivity of the contribution functions to temperature. 
With observations across the prominence, optically thick radiation may be used to infer the temperature structure of the prominence. 
This will however be affected by fluctuations in local density and ionization fraction. 
Multiple observations at different optically thick wavelengths could potentially yield much desired information on the internal temperature of solar prominences. 
However, from the models presented in this study with radii of 1000~km (Table~\ref{table:pmodels}), bands 3 and 6 only produce peak optical thicknesses greater than unity for models with pressures greater than  0.1 and 0.5~dyn cm$^{-2}$, respectively (Fig.~\ref{fig:tau_lambda}). 
%Simultaneous band 3 and band 6 observations were available in cycle 4 or cycle 5, however even if they were it would be unlikely unless very high pressure prominences are observed that both bands would produce optical thick radiation. 
Multiple wavelength observation within the individual channels of either band may be possible, however their close wavelengths may cause them to represent locations of similar kinetic temperature. 
Future ALMA cycles will provide simultaneous and co-spatial observations with more band options available to solar observers.
This shall allow a better understanding into the detailed temperature structure within prominences, as longer wavelength observations will more consistently correspond to optically thick radiation for larger sections of material.

\section{Conclusions}\label{sec:conclusions}

In this study we have modelled brightness temperatures of solar prominences in the wavelength range of ALMA.
We considered a two dimensional cylindrical structure filled with hydrogen and helium, where the ionization and level populations has been calculated under non-LTE conditions. 
Two sets of prominence models were used: isobaric, isothermal small-scale structures, and large-scale structures with radially increasing temperature distributions. 
In both cases, we believe that ALMA shall be able to provide strong thermal diagnostic capabilities, provided that the interpretation of observations is supported by the use of non-LTE simulation results. 

The optically thick radiation in either set of models yields a brightness temperature representative of the kinetic temperature at the region of highest contribution function, located near the point in the LOS where $\tau = 1$. 
In an isothermal model this will be the kinetic temperature across the entire thread. 
Optically thin emission from a non-isothermal thread will present an integration over any thermal structure. 
Multiple observations of optically thin or thick emission for an approximately isothermal thread may be used in conjunction with our models to set constraints on a representative isothermal temperature and isobaric pressure. 

The brightness temperature ratio for two distinct wavelengths also provides an estimate of the respective optical thickness of the structure, as well as the LOS-averaged emission measure. 
Currently, ALMA  only offers non-simultaneous observations in bands 3 and 6 for solar observations. 
In the future, simultaneous observations with a larger range of available bands will be invaluable to progress in our understanding of the conditions within the solar prominence plasma.

%%%%%%%%%%%%%%%%%%%%%%%%%%%%%%%%%%%%%%%%%%%%%%%%%%%%%%%%%%%%%%%%%%%%%%%%%%%
%% Appendix
%
% \appendix   

%%%%%%%%%%%%%%%%%%%%%%%%%%%%%%%%%%%%%%%%%%%%%%%%%%%%%%%%%%%%%%%%%%%%%%%%%%%
%% Acknowledgements
%
\begin{acks}
  AR acknowledges support from a STFC studentship.
  NL acknowledges support from STFC grant  ST/L000741/1.
  This research made use of NASA's Astrophysics Data System.

\end{acks}

%%% %%%%%%%%%%%%%%%%%%%%%%%%%%%%%%%%%%%%%%%%%%%%%%%%%%%%%%%%%%%
%% Bibliography
%
% Using BibTeX
%
\bibliographystyle{spr-mp-sola}
\bibliography{paper}  
%
% Without BibTeX 
% \begin{thebibliography}{}
% \bibitem[\protect\citeauthoryear{Author}{Year}]{key}
%   <bibliographical entry>
%
% \bibitem[\protect\citeauthoryear{}{}]{}
%   
%  
% \end{thebibliography}

\end{article} 
\end{document}